# Origin of Pseudogap and Stripe Phase in High-$T_c$ Superconductors in Two Dimensional Picture

J. K. Srivastava*

*Tata Institute of Fundamental Research, Mumbai-400005, India*

(29 March 2005; revised (v4) 19 August 2005)

The details of the pseudogap origin and other gap related properties discussed earlier for cuprates, in the framework of the paired cluster (PC) model, using three dimensional (3D) electronic density of states (DOS) [1-3], are shown to remain valid even when a two dimensional (2D) cuprate electronic DOS is used. Similarly the stripe phase description is also shown to be similar for the 3D and 2D cases. These results confirm the PC model's capability in explaining the high-$T_c$ superconductivity properties.

## I. INTRODUCTION

In an earlier work [1-3] we have explained the origin of pseudogap and stripe phase in magnetically frustrated cuprate superconductors assuming a three dimensional (3D) electronic density of states (DOS), $D(E_{el})$, for conducting electrons (CEs) and using a paired cluster (PC) model developed by us to describe the physics of high-$T_c$ superconductivity (HTSC); $E_{el}$ = CE energy, $T_c$ = critical temperature, $D_t(E_{el})$, total electronic DOS (at $E_{el}$), $= D_f(E_{el})$ ( filled electronic DOS) $+ D_e(E_{el})$ ( empty electronic DOS ), $D_t(E_{el}) = A E_{el}^{1/2}$ for $T \geq T_c$ and for $T < T_c$,

$$D_t(E_{el}) = A \, \mathrm{Re}[(\, E_F \pm \sqrt{[(E_{el}+i\Gamma)-E_F]^2 - \Delta^2}\, )^{1/2} \{\frac{|(E_{el}+i\Gamma)-E_F|}{\sqrt{[(E_{el}+i\Gamma)-E_F]^2 - \Delta^2}}\}]\quad ,$$

where - (minus) sign before the square root, in the first round bracket, applies for $E_{el} \leq E_F$ and + sign for $E_{el} > E_F$, Re means real part, A= proportionality constant, T = working temperature, $i = (-1)^{1/2}$, $\Delta$ = BCS ( superconducting state (SS) ) energy gap/2, $E_F$ = Fermi energy and $\Gamma$, explained later, = broadening due to Cooper pair (CP) lifetime decay. The assumption of 3D $D(E_{el})$ looks justified due to the following reasons. The conductivity ($\sigma$), and the superconductivity, are 3D in nature in cuprates. Eventhough the c-axis resistivity ($\rho_c$) is higher than a-,b-axis resistivity ($\rho_a$, $\rho_b$ ), it is still in the mΩ-cm range [4, 5], which is similar to $\rho$ (resistivity ) observed in some metals like Au-Fe spin glass[6]. $\rho_c$ is only about an order of magnitude higher than $\rho_a$, $\rho_b$ (due to reasons described in [1-3]) and thus the system (cuprate) is not an insulator along the c-axis; $\rho$ for an insulator is in MΩ-cm range [6, 7]. Also careful measurements show same $T_c$ along the a-,b-,c-axes [4, 5]. A mΩ-cm range $\rho$ is observed in powdered samples (powdered cuprates) also, where a-,b-,c-axes are random [4, 5]. If $\rho$ was insulating along any direction, powder sample would have shown insulating $\rho$ (MΩ-cm range value). Also experiments show same $\rho$ mechanism for $\rho_a$, $\rho_b$, $\rho_c$ below, and above, $T_c$ [1, 4, 5]. Thus the use of 3D $D(E_{el})$ looks justified. However at the same time it is also true that the cuprates have layered (two dimensional (2D)) structure where Cu-O planes are a-,b-planes, stacked one over the other along the c-axis in the crystal unit cell [1, 4, 5]. It is the coupling between the Cu-O planes which gives three dimensionality, but this coupling has different strength in different cuprates. Thus some 2D nature (two dimensionality) is always present in cuprates [4, 5]. In some systems, like Bi-cuprates [5, 8], this two dimensionality may be more pronounced. It is thus necessary to ascertain that the results obtained earlier [1-3] using 3D $D(E_{el})$ are valid even for 2D cuprate $D(E_{el})$. In this paper we examine this point and find that the 2D $D(E_{el})$ based results are same as those obtained earlier [1-3] using 3D $D(E_{el})$. In the following sections we give the details.

## II. METHODOLOGY AND RESULTS

Since band structure calculation conduction bands for 3D lattices can be approximated to a parabolic shape, quadratic approximation DOS can be used there reasonably well [1, 4, 5, 9, 10]. However this is not the case for a 2D cuprate lattice (Cu-O planar structure with small plane-plane coupling) owing to the presence of van Hove singularity in their electronic DOS (conduction band) at $E_{el}=E_s$ where $E_s$ (the singularity point energy) is close to $E_F$ [4, 5, 8, 11-16]. Several people have tried to associate the cuprate high-$T_c$ with the presence of this singularity [4, 5, 12]. However since experimentally such a singularity has not been observed, it is believed that any small coupling between the Cu-O planes, which can always be present, turns this singularity into a broad peak at $E_s$ and therefore the broadening, $\Lambda$, should be taken into account in any calculation [13, 14]. Thus for the 2D Cu-O planes [4, 5, 8, 12-19],

$$D_t(E_{el}) = C[1 + \frac{1}{2} \ln \frac{E_s}{\sqrt{(E_{el}-E_s)^2 + \Lambda^2}}] \quad (1)$$

for $T \geq T_c$, and for $T < T_c$,

$$D_t(E_{el}) = C \operatorname{Re}[\{1 + \frac{1}{2} \ln \frac{E_s}{\sqrt{([\{(E_{el}+i\Gamma)-E_F\}^2 - \Delta^2]^{1/2} + E_F - E_s)^2 + \Lambda^2}}\}\{\frac{|(E_{el}+i\Gamma)-E_F|}{\sqrt{[(E_{el}+i\Gamma)-E_F]^2 - \Delta^2}}\}], \quad (2)$$

where C = proportionality constant and $\Gamma$, as mentioned before, is the broadening present owing to the CP lifetime decay arising due to the inelastic electron- electron and electron- phonon scatterings [5, 17-19]. We examine below the effect of this (Eq. 1, 2) 2D $D(E_{el})$ on the pseudogap and stripe phase in cuprates. For the pseudogap calculation, we discuss the $T \geq T_c$ and $T < T_c$ ranges separately and assume a small plane- plane coupling ($\Lambda/E_s = 5\%$) which is sufficient to give a broad singularity peak.

### (i) $T_c \leq T \leq T_{CF}$

Earlier [1-3], for the 3D $D(E_{el})$ case, we have presented $D(E_{el})$ vs. $E_{el}$ variation for $D_t(E_{el})$, $D_f(E_{el})$ and $D_{fr}(E_{el})$, the density of filled states redistributed. $D_f(E_{el}) = D_t(E_{el}) \times f(E_{el})$, where $f(E_{el})$ is the Fermi function, and $D_{fr}(E_{el})$ is the modified $D_f(E_{el})$ i.e. the $D_f(E_{el})$ which has got modified owing to the $\Delta E_{el}$ scattering which occurs for $T \geq T_c$ in the PC model [1-3]. According to the PC model [1-3], paired magnetic clusters are present in the cuprate lattice below a temperature $T_{CF}$ (> $T_c$); $T_{CF}$ = cluster formation temperature (pseudogap appearance temperature) [1-3]. The CEs for $T \geq T_c$, and both the CEs and the Cooper pairs, CPs, for $T < T_c$, interact with these clusters, by an interaction described in [1-3], and this interaction enhances the CEs' energy, $E_{el}$, by $\Delta E_{el}$ and CPs' energy, $E_{CP}$, by $\Delta E_{CP}$. The modification in $D_f(E_{el})$ due to the $\Delta E_{el}$ enhancement for $T \geq T_c$, and due to both the $\Delta E_{el}$ and $\Delta E_{CP}$ enhancements below $T_c$, gives rise to $D_{fr}(E_{el})$. The earlier calculations [1, 3] have been presented for some typical values of the parameters used (i.e. T, $E_F$, $\Delta E_{el}$, $N_P$ where $N_P \equiv (N_P)_{CE}$, is the percentage of CEs for which $\Delta E_{el}$ enhancement occurs and its value depends on the relative space occupied by the clusters and the cluster boundaries in the cuprate lattice [1-3]; below $T_c$ we have $\Delta E_{CP}$ and $(N_P)_{CP}$ (percentage of CPs for which $\Delta E_{CP}$ enhancement occurs) also present). As has been explained there [1, 3], though the results are presented for certain typical parameter values they have been checked to be general in nature. The typical parameter values have been chosen for obtaining results similar to those of the tunneling conductance experiments mentioned in [1, 3], or to bring out any specific nature of the $D_{fr}(E_{el})$, or $D_{er}(E_{el})$ [redistributed $D_e(E_{el})=D_t(E_{el})-D_{fr}(E_{el})$], vs. $E_{el}$ curve, and are consistent with the theoretical estimates [1-3].

For the present 2D lattice, $D_t(E_{el})$ is given by Eq. (1) for $T \geq T_c$ and the other quantities are as follows.

$$D_f(E_{el}) = D_t(E_{el}) \times f(E_{el}), \quad (3)$$

where $f(E_{el}) = 1/\{\exp[(E_{el} - E_F)/k_B T]+1\}$ and $k_B$ = Boltzmann's constant. Assuming that the Pauli principle permits the above mentioned $\Delta E_{el}$ scattering, i.e. empty states are available for such a scattering, we have [1, 3],

$$D_{fr}(E_{el}) = D_f(E_{el}) - N_P D_f(E_{el}) + N_P D_f(E_{el} - \Delta E_{el}). \tag{4}$$

Details for the occurrence of this scattering for any $E_{el}$ are discussed in [1, 3]. Fig. 1 shows the results obtained, where $E_{el}$ dependence of $D_t(E_{el})$ (dotted curve), $D_f(E_{el})$ (dashed curve) and $D_{fr}(E_{el})$ (full line and dash- dot (a, b) curves) are plotted. The parameter values chosen are the same as used in [1, 3] to facilitate comparison except $E_s$, $\Lambda$ which are new and, as mentioned before, have been fixed to $\Lambda/E_s = 5\%$; T=200K, $E_F = E_s = 310$ meV, $\Lambda = 15$meV, $\Delta E_{el}(E_F) = 300$ meV, $N_P = 50\%$ (full line curve), 40% (curve a) and 60% (curve b), and $\alpha = 0.7$ (meV)$^{-1}$, where $\Delta E_{el}(E_F)$ is the value of $\Delta E_{el}$ at $E_F$ and as discussed in [1, 3] $\alpha$ describes the dependence of $\Delta E_{el}$ on $E_{el}$. The results obtained in Fig. 1 are same as those obtained in [1, 3]. The Fig. 1 results have been checked to be valid for $E_s$ close to $E_F$, either $> E_F$ or $< E_F$, cases also. Thus the results which have been obtained for a 3D $D(E_{el})$ in [1, 3] for $T \sim T_{CF}$ are valid for 2D cuprate $D(E_{el})$ also. Similarly the other results obtained in [1, 3] for $T \geq T_c$ have also been found to exist in the present 2D $D(E_{el})$ case. Further, as discussed in [1, 3], even if CPs' presence is assumed in $T_c \leq T \leq T_{CF}$ range, results obtained here remain same.

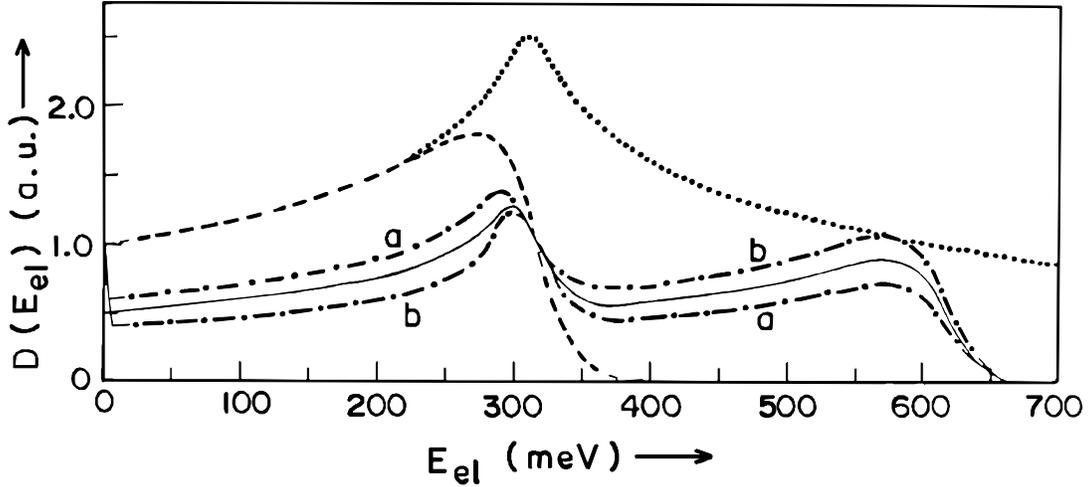

Fig.1. Dependence of the electronic density of states, $D(E_{el})$, on electrons' energy, $E_{el}$, for $T > T_c$; $T_c$= critical temperature, a.u.= arbitrary unit. Details are described in the text.

**(ii) $0 \leq T < T_c$**

Below $T_c$, BCS energy gap is also present alongwith the $\Delta E_{el}$ and $\Delta E_{CP}$ scatterings in the $D(E_{el})$ vs. $E_{el}$ distribution [1-3]. In this case $D_t(E_{el})$ is given by Eq.(2), $D_f(E_{el})$ by Eq.(3) and $D_e(E_{el}) = D_t(E_{el}) - D_f(E_{el})$. To facilitate comparison we have used the parameters of 3D $D(E_{el})$ calculation case [1, 3] and done calculation for the present 2D $D(E_{el})$ case. Fig. 2 shows a typical result where $D_t(E_{el})$ (dotted curve), $D_f(E_{el})$ (dashed curve), $D_{fr}(E_{el})$ (full line curve) and $D_{er}(E_{el})$ (dash - dot curve) vs. $E_{el}$ is shown. The parameters used are T = 4.2K, $E_F = E_s = 310$ meV, $\Lambda = 15$ meV ($\Lambda/E_s = 5\%$), $\Delta = 45$ meV, $\Gamma = 2.25$ meV ($\Gamma/\Delta = 5\%$), $\Delta E_{CP} \sim 0$ (< 1 meV), $\Delta E_{el}(E_F) = 90$ meV, $N_P = 20\%$ and $\alpha = 0.7$ (meV)$^{-1}$. The dash - double dot curve (curve b) is the total electronic DOS curve which would have existed at 4.2K if there was no BCS gap present (i.e. $D_t(E_{el})$ of Eq.(1)).To distinguish it from the $D_t(E_{el})$ of Eq.(2) (dotted curve in Fig. 2), we denote it (curve b) by $D_t'(E_{el})$. Since below the gap ($E_{el} < (E_F - \Delta)$), $f(E_{el}) = 1$ at 4.2K, curve b also represents the filled electronic DOS which would have existed at 4.2K, for $E_{el} < (E_F - \Delta)$, if no BCS gap was present, i.e. $D_f'(E_{el})$. This means below the gap $D_t'(E_{el}) = D_f'(E_{el})$. Thus $D_f'(E_{el})$ gives the filled DOS at 4.2K for those CEs which have not formed CPs and $D_f(E_{el}) - D_f'(E_{el})$ for those which have formed CPs. Since, as mentioned above in the parameter values, $\Delta E_{CP} \sim 0$ at 4.2K, we have for the region below the gap $(E_{el} < (E_F - \Delta))$ [1, 3],

$$D_{fr}(E_{el}) = D_f(E_{el}) - N_P\, D_f'(E_{el}), \qquad (5)$$

and for the regions inside and above the gap,

$$D_{fr}(E_{el}) = N_P\, D_f'(E_{el} - \Delta E_{el}). \qquad (6)$$

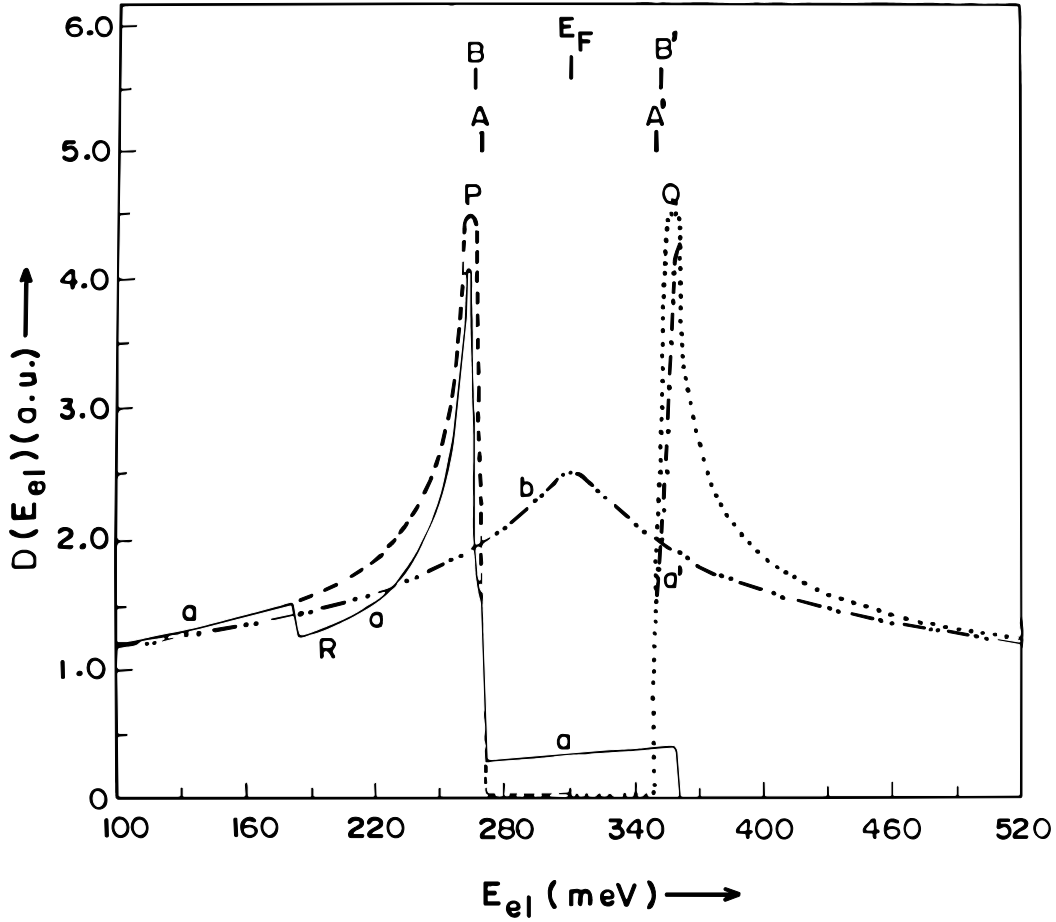

Fig.2. Dependence of the electronic density of states, $D(E_{el})$, on electrons' energy, $E_{el}$, for $T < T_c$; $T_c$ = critical temperature, a.u.= arbitrary unit. Details are described in the text.

In Eq.(5), the term $N_P\, D_f'(E_{el} - \Delta E_{el})$ is absent on the right hand side (R.H.S.) since all the $D_t'(E_{el})$ states below the gap are completely filled at 4.2K ($D_t'(E_{el}) = D_f'(E_{el})$). In Eq.(6) only one term appears on R.H.S. since inside and above the gap there exist negligible, or no, filled states in the absence of $\Delta E_{el}$ scattering. However if any significant number of filled states are present in those regions in some case, then they will get added to the R.H.S. of Eq.(6) (Appendix).

The results obtained in Fig. 2, like change in the size, shape, location of BCS peaks (P, Q), appearance of dip R, enhancement of A, A′ separation to B, B′ separation, etc., due to the $\Delta E_{el}$ scattering presence are same as obtained for the 3D $D(E_{el})$ case in [1, 3]; A, A′ are the points where the curve b intersects the gap edges when no $\Delta E_{el}$ scattering is present and B, B′ when $\Delta E_{el}$ scattering exists [1, 3].

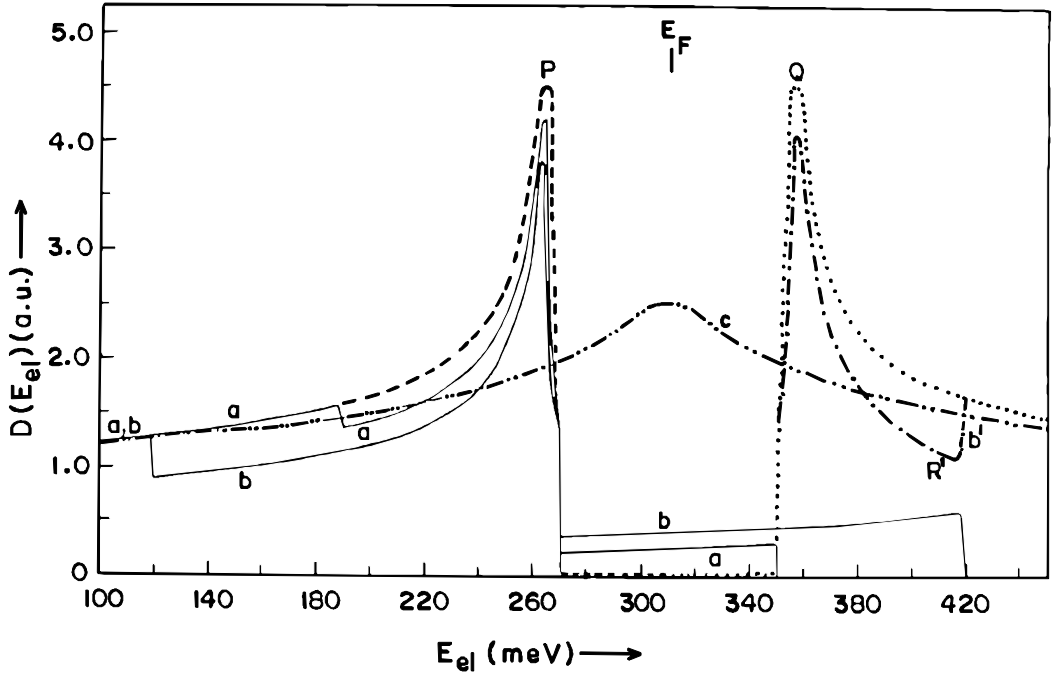

Fig.3. Dependence of the electronic density of states, $D(E_{el})$, on electrons' energy, $E_{el}$, for $T < T_c$ and some parameter values different from the Fig. 2 parameter values; $T_c$= critical temperature, a.u.= arbitrary unit. Details are described in the text.

The Fig. 3 shows another typical result. Here the dotted, dashed, dash - dot, dash - double dot and full line curves have the same meanings as in Fig. 2 and all the parameter values are same as those of Fig. 2 except $N_P = 15\%$, $\Delta E_{el}(E_F) = 80$ meV for the curve a and $N_P = 30\%$, $\Delta E_{el}(E_F) = 150$ meV for the curve b. A comparison of Fig. 3 results with those obtained in [1, 3] shows that they are the same. Similarly for the other situations discussed in [1, 3], like $T \sim T_c$ where $\Delta E_{CP}$ scattering too is present (Appendix), also we get same results for the 3D $D(E_{el})$ and 2D $D(E_{el})$ cases. Thus the conclusions drawn on the basis of the 3D $D(E_{el})$ calculation results [1, 3] are valid for the present 2D $D(E_{el})$ calculations also.

**(iii) Stripe phase**

For the 3D case, the stripe phase is discussed in [1, 3] and as mentioned there the 3D stripes will have tubular structure with 2D planar projection on Cu-O planes. However for the 2D case, the stripes will be located in Cu-O planes only. The other stripe properties, like bending of stripes, stripe fluctuation etc. [1, 3], remain same for the 3D and 2D cases. Like 3D stripes, the 2D stripes will also have a width and run randomly in the Cu-O plane changing their location with time as the $Cu^{2+} \leftrightarrow Cu^{3+}$ fluctuation [1-3] occurs in superconducting cuprates. However it may be noted that, as discussed in [1, 3], these stripes are formed by the cluster boundaries (hole rich) and are not charge ordered stripes as seen in insulators with complete separation of hole full and holeless regions. Theoretically also such static charge ordered stripes can not be formed in superconducting cuprates [20] due to the presence of next nearest neighbour hopping effect, which is needed for conductivity (metallicity), and so superconductivity, but which suppresses the static stripe formation (charge ordering).

## III. CONCLUSION

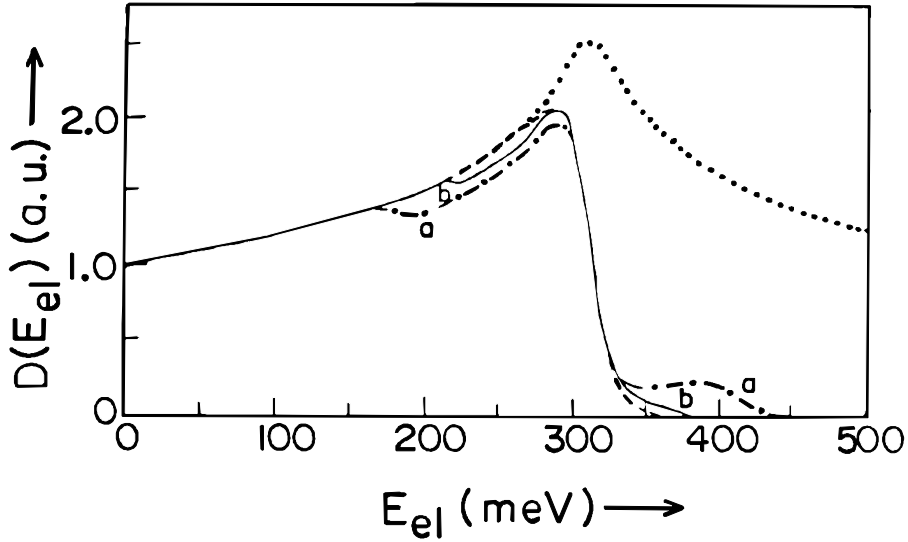

Fig.4. Dependence of the electronic density of states, $D(E_{el})$, on electrons' energy, $E_{el}$, for $T>T_c$ and dopant concentration > critical dopant concentration; $T_c$=critical temperature, a.u.= arbitrary unit. Details are described in the text.

In conclusion, the results obtained for the 3D $D(E_{el})$ case [1, 3] are valid for the 2D $D(E_{el})$ case also. This is true even for those results which are not specifically described above. For example, as has been discussed in [1, 3] above a certain critical dopant concentration, in the overdoped region, pseudogap does not occur in cuprates. This absence has been explained there [1, 3] on the basis of the decrease in $\Delta E_{el}$, $N_P$ values at higher dopant concentrations. In Fig. 4 we have repeated the 3D $D(E_{el})$ calculations for the 2D $D(E_{el})$ case using the same parameter values as those of 3D case [1, 3]; T = 100K, $E_s = E_F$ =310 meV, $\Lambda$ = 15 meV, $\alpha$ = 0.7 (meV)$^{-1}$ and, $\Delta E_{el}(E_F)$ = 100 meV, $N_P$ = 10% for the curve a and $\Delta E_{el}(E_F)$ = 50 meV, $N_P$ = 5% for the curve b. A comparison of the Fig. 4 results with those of 3D $D(E_{el})$ results [1, 3] shows that they are the same. Similar is the case with the other results also, obtained in [1, 3] for the $T \geq T_c$ or $T < T_c$. Thus the results of the 3D case, and the conclusions drawn from them, are valid for the 2D case also. It may be mentioned here that the present results, and so also those of [1, 3], are consistent with some recent experiments which link the pseudogap origin to the change in the electronic DOS near $E_F$ and indicate that the underdoped and overdoped cuprates have the same superconductivity origin [21]. Similarly the $\Delta\theta_D$ break at $T_{CF}$, mentioned in [1-3] ($\theta_D$ = Debye temperature), can be attributed only to the presence of clusters, and cluster boundaries, interacting with CEs [1-3], since both the $\Delta\theta_D$ break and the CE- cluster, - cluster boundary, interaction are present in superconducting cuprates only whereas the regular charge ordered stripe phase (resulting from the absence of $Cu^{2+}\leftrightarrow Cu^{3+}$ type charge fluctuation) is present in nonsuperconducting cuprates where no $\Delta\theta_D$ break is observed [1]. It may be mentioned here that same results are obtained if extended van Hove singularity is used in the DOS calculation instead of logarithmic singularity. Thus the PC model is capable of explaining the HTSC properties. Predictions of the model [1-3] made some years ago, like the coexistence of pseudogap and superconducting state energy gap below $T_c$ or the presence of spin glass (SG) interactions and clusters in high-$T_c$ systems, are being found true by the experiments now [22, 23] confirming the correctness of the model.

## ACKNOWLEDGEMENTS

Useful discussions with V. R. Marathe (TIFR, Mumbai) and S. M. Rao (National Tsing Hua Univ., Taiwan) are greatefully acknowledged.

## APPENDIX

For T ~ $T_c$ case, where $\Delta E_{CP}$ scattering is also present [1, 3], we have for the region below the gap ($E_{el} < (E_F - \Delta)$),

$$D_{fr}(E_{el}) = D_f(E_{el}) - N_P D_f{'}(E_{el}) + N_P D_f{'}(E_{el} - \Delta E_{el}) - (N_P)_{CP} [D_f(E_{el}) - D_f{'}(E_{el})] + (N_P)_{CP} \times$$

$$[D_f(E_{el} - \Delta E_{el}) - D_f{'}(E_{el} - \Delta E_{el})]. \qquad (i)$$

The third term of the above equation will be nonzero only in very few cases when some empty states are present in $D_f{'}(E_{el})$ below the gap, like, for example, Fig. 8(a) case of the first reference of [1] where the temperature spread of $D_f(E_{el})$ curve's tail portion is slightly greater than the gap value. Similarly for the region inside the gap we have,

$$D_{fr}(E_{el}) = N_P D_f{'}(E_{el} - \Delta E_{el}) + D_f(E_{el}). \qquad (ii)$$

The second term in the above equation contributes only if there are some filled states in the gap; generally this is zero or very small. Finally, for the region above the gap,

$$D_{fr}(E_{el}) = N_P D_f{'}(E_{el} - \Delta E_{el}) + D_f(E_{el}) + (N_P)_{CP} [D_f(E_{el} - \Delta E_{el}) - D_f{'}(E_{el} - \Delta E_{el})]. \qquad (iii)$$

As in Eq.(ii), the second term of Eq.(iii) is zero or very small. The third term of Eq.(iii) contributes only if some excited CPs are present above the gap.

---